\documentclass[sigconf]{acmart}
\begin{document}

\title{NLDS-QL: From natural language data science questions to  queries on graphs: analysing patients conditions \& treatments}

\author{Genoveva Vargas-Solar}
\orcid{1234-5678-9012}
\affiliation{%
  \institution{French Council of Scientific Research (CNRS), LIRIS}
  \city{Lyon} 
  \postcode{38000}
    \country{France}
}
\email{genoveva.vargas-solar@cnrs.fr}

\author{Karim Dao}
\affiliation{%
  \institution{University Paris Dauphine}
  %\streetaddress{F-69622}
  \city{Tunis} 
  \country{Tunisia} 
}
\email{karim.dao@dauphine.tn}

\author{Mirian Halfeld-Ferrari}
\affiliation{%
  \institution{Université d'Orléans, INSA-CVL LIFO}
  \city{Orléans} 
  \country{France}}
\email{mirian@univ-orleans.fr}

\begin{abstract}
  This paper introduces NLDS-QL \footnote{This work was partially funded by the French CNRS MADICS action DOING and the UEMOA (Union économique et monétaire ouest-africaine) funding K. Dao's master.}, a translator of data science  questions expressed in natural language (NL) into data science queries on graph databases.   Our translator is based on a simplified NL described by a grammar that specifies sentences combining  keywords to refer to operations on graphs with the vocabulary of the graph schema.  The demonstration proposed in this paper shows NLDS-QL in action within a scenario to explore and analyse  a graph base with patient diagnoses generated with the open-source Synthea.
\end{abstract}

\maketitle



%________________________________________________________
\section{Introduction}
%________________________________________________________

The volume of connected data, often modelled as graphs, has grown exponentially. The availability of these graphs data collections has been democratised through social networks and knowledge graphs used to explore content (e.g. scientific papers, clinical cases).  
Although this accessibility is promising, it introduces a barrier for non-experts, who have to familiarise with the nature of the data, the way they have been represented in the database and the specific query languages or user interfaces to access them.

Besides, the emergence of data science has brought a new type of 'complex' queries embodying a data analysis scenario. 
A data science query generally refers to a workflow of tasks including exploration, data cleaning and preparation, sampling and analysis. These workflows include visualisation and evaluation tasks that involve the calculation of scores and metrics.  Implementing these workflows is a challenge even for engineers and data scientists. In most cases,  users should have advanced skills in querying and analysing the data according to their needs and the type of search questions to be answered. Requiring formal and technical expertise from non-computer scientists is not obvious or reasonable.

In our work, the goal is to identify important research questions,
formulated by non-technical experts over data collections.
Allowing their expression in natural language (NL) would optimize data accessibility.
However, this facility implies complex NL analysis, particularly when such questions are
intended to be transformed into sophisticated workflows.
Thus, to achieve our goal, we address the problem under a
reverse-engineering strategy: we build a simplified natural language (NL) grammar and map
NL data science questions to graph data science queries as those proposed by Neo4J DS templates.
The users can then express their questions using this simplified NL with sentences that correspond
to the DS Neo4J templates, seeing and \textit{assessing the results and proposing new questions over an
enriched vocabulary that can extend the NL query language treated by our tool.}
%%%
%}

%----------------------------------------------------
\paragraph{Contribution}
%----------------------------------------------------

We propose NLDS-QL,  a semi-automatic and evolutive interface for processing experts' NL questions on a given vocabulary. It derives data science query templates that program the answers to these questions, and offers a conversational evaluation of results and adapted vocabulary. NLDS-QL is based on a simplified English NL  that associates a vocabulary based on graph schema keywords to refer to operations on graphs (e.g.,  attributes describing the nodes, links, and associated labels). Depending on the type of queries to explore graphs or to analyse them, their expression in NL can yield to a  more or less complex document. 
%}

%----------------------------------------------------
\paragraph{Demonstration}
%----------------------------------------------------
NLDS-QL will be presented in the context of medical diagnostics using the patient graph of the Synthea \footnote{ https://xilinx.github.io/graphanalytics/recom-tg3/synthea-overview.html
} data collection and using Neo4J for running translated queries. Assuming that they have a collection of data corresponding to medical follow-ups, doctors (users)  express the questions whose answers would be helpful for the elaboration of a diagnosis and their decision making. Questions described in written or spoken English can denote navigational, aggregation and data science queries requiring centrality and clustering algorithms to be expressed and answered. Given the ambiguity of the NL, the translation of questions can lead to several Cypher (data science) queries. So the demonstration of NLDS-QL is proposed under a conversational pipeline where users acquainted with Cypher can choose the query that best corresponds to their expectations before executing them. Non-expert users can decide to let the system run different Cypher queries, analyse their results and then choose one or adjust their question.

The remainder of the paper is organised as follows. Section \ref{sec:related} synthesises the main families of works addressing NL to query languages translation and processing. 
Section \ref{sec:architecture} describes the general architecture of NLDS-QL.    Section \ref{sec:demonstration} describes the demonstration scenario for NLDS-QL and the tasks proposed to users during the demo.
Section \ref{sec:conclusion} concludes the paper and discusses lessons learned from the demonstration.
%\igvs{Here a sentence describing the demo idea}

%----------------------------------------------------
\section{Related work}\label{sec:related}
%----------------------------------------------------

Existing work has addressed data mining using NL, but not yet the expression of data science questions.
There are different approaches to developing a NL interface for database queries, in general for relational systems. Georgia Kutrika \cite{koutrika2021}  describes the process of handling NL queries with a workflow that consists of three steps. Given a relational database and assuming knowledge of the schema vocabulary:
1. Analysis of the NL query expression;
2. Disambiguation and interpretation, which produces a set of ranked interpretations;
3. Finally, the translation into SQL and its execution.
Three generations of NL Query to SQL Transformation systems can be identified \cite{gkini2021depth}, namely: 
%\begin{itemize}
   % \item 
   %\noindent
   (i) Keyword based i.e. information retrieval techniques to evaluate queries. For example, systems like Discover (query interpretations as subgraphs), DiscoverIR \cite{song2015tr,lakhanpal2015discover}, Spark (ranking and fast execution) \cite{thomas2020natural}.
    %\\
    %\item 
  %   \noindent
   (ii) NL processing based like NaLIR (parser) \cite{li2014nalir}, ATHENA \cite{saha2016athena} (ontologies and mappings).
   %\\
%    \item 
%\noindent
   (iii) Machine translation using neural networks  \cite{baevski2020wav2vec}, like NL to SQL conversion as a language translation problem. The challenge is  
training a neural network on a large number of {NL/SQL query} pairs. Approaches like SQLNet \cite{xu2017sqlnet}, Hydranet adopt this strategy.
%\end{itemize}

% Two families of works can be identified concerning Graph Querying in Natural Language.
% 1. Approaches that address natural language translation on SPARQL for knowledge graph queries. For example, Machine learning techniques (Tree-LSTM and neural networks)  \cite{baevski2020wav2vec} and
% methods based on grammar and logical predicates.
% 2. Approaches that address the creation of structured queries using natural language processing methods such as named entity recognition. For example, recognising named entities and binary relationships (pattern extraction); identifying key entities and relationships corresponding to graph components. 

% \todo[inline]{Mi: I do not know if I understand the next paragraph correctly. I propose a new version - but please check if my interpretation is good, specially for item 2, which is not that clear for me. You mean... the query is kept in NL ?\\
%  \vspace{1cm}

On the other hand, we identify two families of works concerning NL Graph Querying.
1. Approaches that address NL translation on SPARQL for knowledge graph queries, such as \cite{baevski2020wav2vec},  concerning machine learning techniques (Tree-LSTM and neural networks),  and methods based on grammar and logical predicates \cite{shiqi2021,liu2018neural}.
%\textcolor{blue}{A reference here???}. 
%\\
2. Approaches that translate questions into structured queries using NL processing methods such as:  named entity recognition, binary relationship (pattern) extraction,  key entity identification, and relationship mapping to graph components \cite{oro2015natural}.
%}
%

The literature agrees that little work has addressed the issue of answer validation, i.e. how can a user confirm that the results match the query's intent? 
With the emergence of data science, two questions arise:
How to express data exploration, cleaning and preparation, sampling and analysis, visualisation and metrics calculation?
How to model and process research questions formulated by non-technical experts? How to allow their NL expression and translation into data science queries?
%
% \todo[inline]{Mi: INSTEAD OF the paragraph below...}

% Thus  our work  proposes an approach to query and analyse 
% graph databases in natural language 
% to facilitate these tasks for experts.

% \todo[inline]{Mi: A NEW PROPOSAL FOR THE END of this section (TO BE ADAPTED):\\
% \vspace{0.3cm}
%
Our work proposes an interactive reverse-engineering method that highlights essential aspects to be considered when dealing with  NL medical queries. It is the basis for a user-friendly  interface adapted for the medical personnel.
%}

%________________________________________________________
\section{NLDS-QL}\label{sec:architecture}
%________________________________________________________

\begin{figure*}[h!t] \centering
\includegraphics[width=0.95\textwidth]{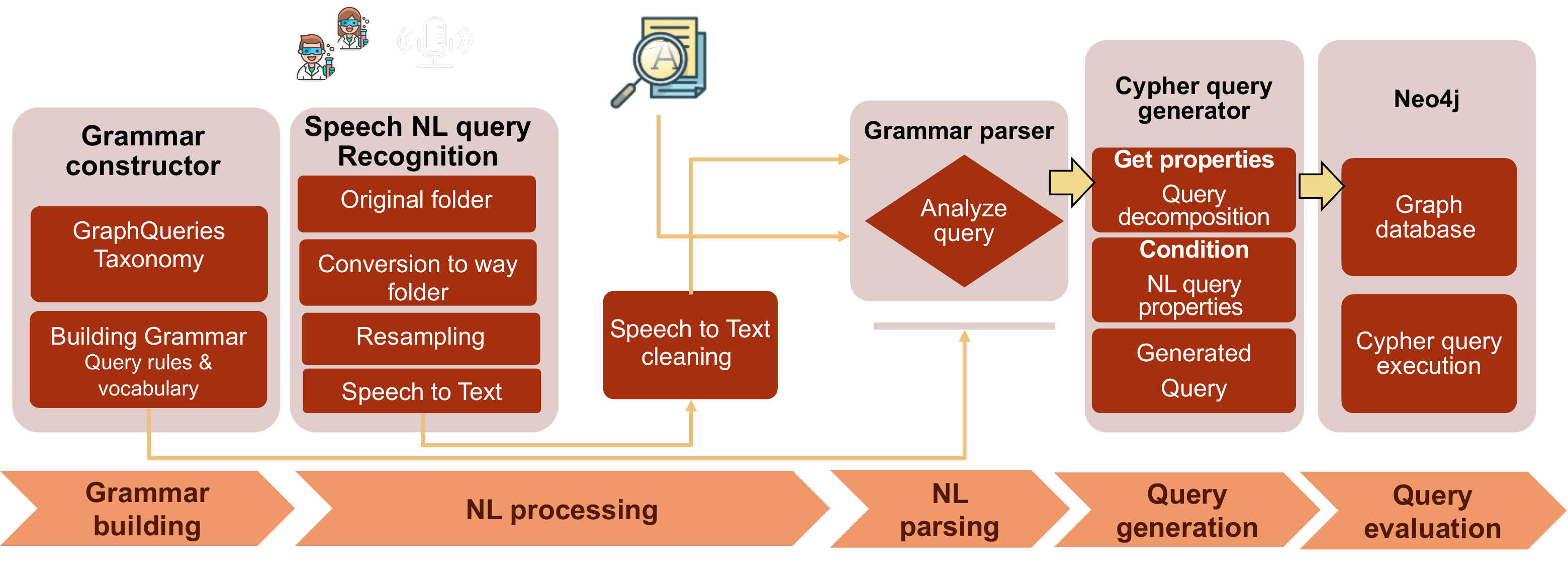}
\caption{NLDS-QL general architecture
}\label{fig:dsnl-ql}
\end{figure*}

The general process implemented by NLDS-QL is shown in Figure \ref{fig:dsnl-ql}.
The first two phases of our translation approach are devoted to analysing the NL query, which is expressed as a text (see the NL processing and NL parsing phases in Figure \ref{fig:dsnl-ql}). The text can be written or defined as a voice message and transcribed into text.
Therefore, the NL processing phases implement the classical text processing of syntactic analysis to produce an expression tree that represents the query (this is done by a parser as shown in Figure \ref{fig:dsnl-ql}).
The tree is then processed to produce one or more corresponding Cypher queries in the query generation phase (see the query generation phase in Figure \ref{fig:dsnl-ql}).
Finally, the queries are evaluated on Neo4J (see the query evaluation phase in Figure \ref{fig:dsnl-ql}).

%---------------------------------------------------------
\paragraph{Overview of NLDS-QL expressions.}
%---------------------------------------------------------
The expression of NLDS-QL questions is based on the way data science operations are applied on graphs in Neo4J. Neo4J defines a general template including several commands for expressing the execution of a DS query.

DS operations are generally applied on graph views created in memory from persistent graphs. The views require main memory space to be allocated for creating them and main memory resources for using algorithms with specific execution conditions expressed in parameters. Thus, Neo4J provides commands for performing these estimations and then calling DS operations with given parameters' values. Finally, DS operations can yield new graphs that can be named and persist or not. The creation of new graphs and whether to persist them is expressed as function call commands. 

 Consequently, the definition of data science questions in NL include expressions for specifying the commands specified in the Neo4J template. 
 The most simple expression for defining a graph view and estimating the memory required assuming that it is stored in Neo4J and that the graph schema with the nodes and relations is available, is defined with the following English expression: 
 
 \noindent
 - {\em Create and estimate memory for the graph view} $<$subgraph-name$>$ [{\em named as} $<$ view name $>$] {\em with the node} $<$ node name $>$ {\em and} {\em the relationship} $<$ relationship name $>$ [{\em oriented}]
 
 The data science operation task includes estimating the cost in memory of applying a graph data science algorithm on the graph, using the algorithm on the graph view. According to specific keywords, NLDS-QL can determine the type of algorithm that can be applied. Keywords like {\em most important, most popular, most influential} refer to centrality algorithms such as PageRank and Louvain and {\em classify, communities, group} can refer to clustering algorithms like Label Propagation, as illustrated by the following three questions:
 
 \noindent
- $Q_1:$ {\em Estimate the required memory for applying} $<$ DS algorithm name$>$ {\em on the graph view} $<$ view name $>$

  \noindent
 - $Q_2:$ {\em Find the most important/most popular} $<$ node name $>$ {\em with}  $<$ relation name $>$ [{\em in the graph } $<$ graph name $>$] {\em with} $<$ number of iterations $>$ {\em maximum of iterations and with a damping factor} $<$ floating number $>$

\noindent
- $Q_3:$ {\em Classify/Find groups/communities of} $<$ node name $>$ {\em within the view}  $<$ view name $>$  {\em with relation} $<$ relation name $>$ {\em with} $<$ number of iterations $>$ {\em maximum of iterations}

%________________________________________________________
\section{Demonstration overview} \label{sec:demonstration}
%________________________________________________________

We set up an experiment to validate our approach. Therefore we use the patient part of the Synthea Generic study. 
The Synthea's Generic Graph \footnote{{https://xilinx.github.io/graphanalytics/recom-tg3/synthea-overview.html
}}
models  various diseases 
conditions that contribute to the medical history of synthetic patients
800K vertices and nearly 2000K edges.

Querying graphs is based on navigational queries, which retrieve information already "contained" in a graph.
In the Synthea graph, it is possible to ask simple queries like :
How many patients are there in the Synthea study? Which allergies are identified in the Synthea study patients? 
But it is also possible to go further and ask analytical type questions that involve classification tasks such as What are the most frequently prescribed drugs for patients in the Synthea study?
Answering this type of query involves performing a sequence of tasks ordered in a pipeline, which we call a data science query.

\begin{figure}[ht!] \centering
\includegraphics[width=0.45\textwidth]{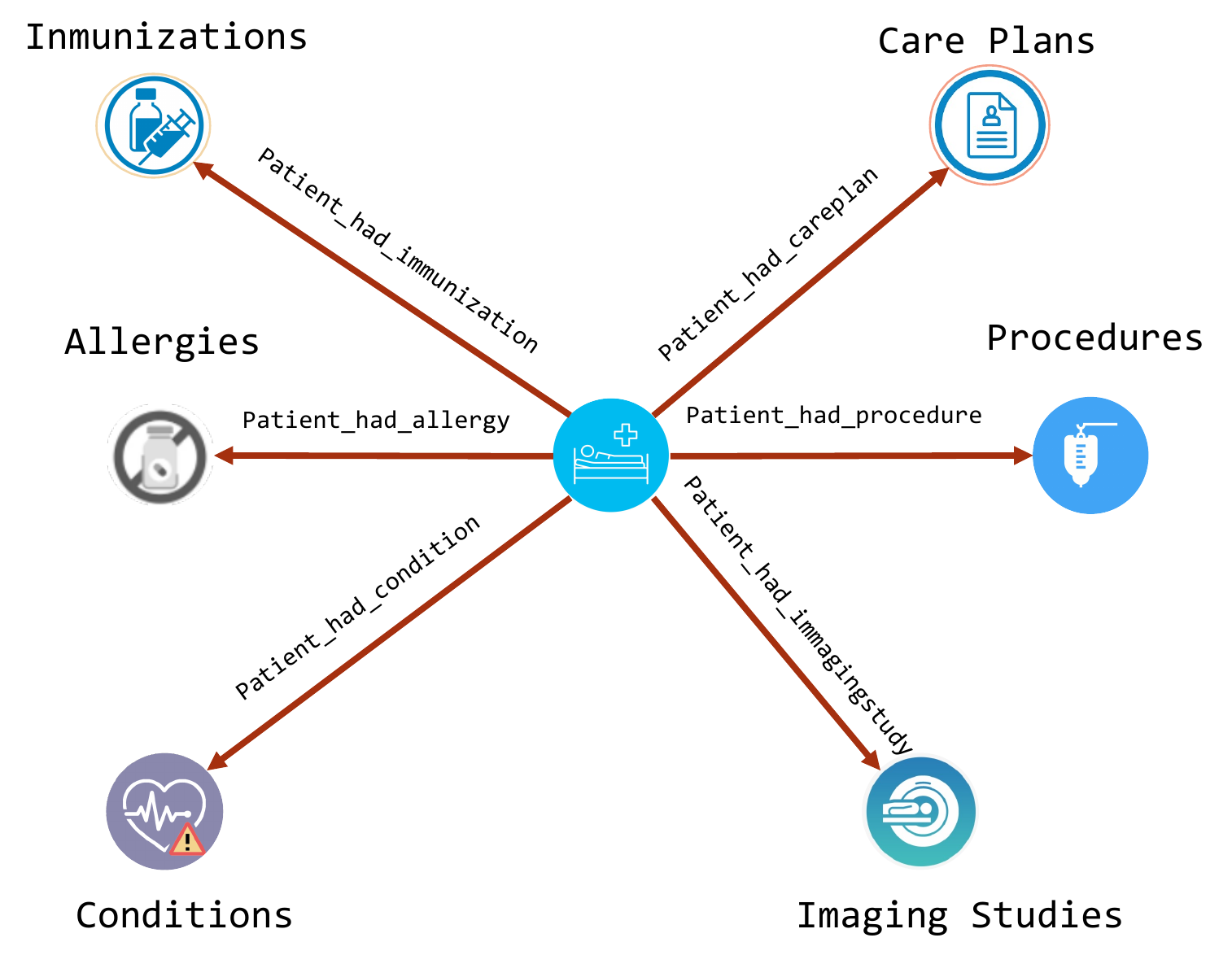}
\caption{Synthea patients' graph
}
\label{fig:demo-setting}
\end{figure}

%.    .    ..    .    ..    .    ..    .   
\paragraph{Demonstration scenario}
%.    .    ..    .    ..    .    ..    .    ..    .    ..    . 
The demonstration is based on the Synthea {\bf patients graph} shown in Figure \ref{fig:demo-setting}. It describes the immunisations, allergies, conditions, studies, procedures and care plans of patients. Each entity and its relations are characterised by properties that describe them.
The patient graph has approximately 100 thousand nodes and 37 thousand relations stored on  Neo4J. The demonstration runs on a local machine to avoid connection problems, and the questions use the graph vocabulary extracted from its schema.

\begin{figure*}[h!t] \centering
\includegraphics[width=0.85\textwidth]{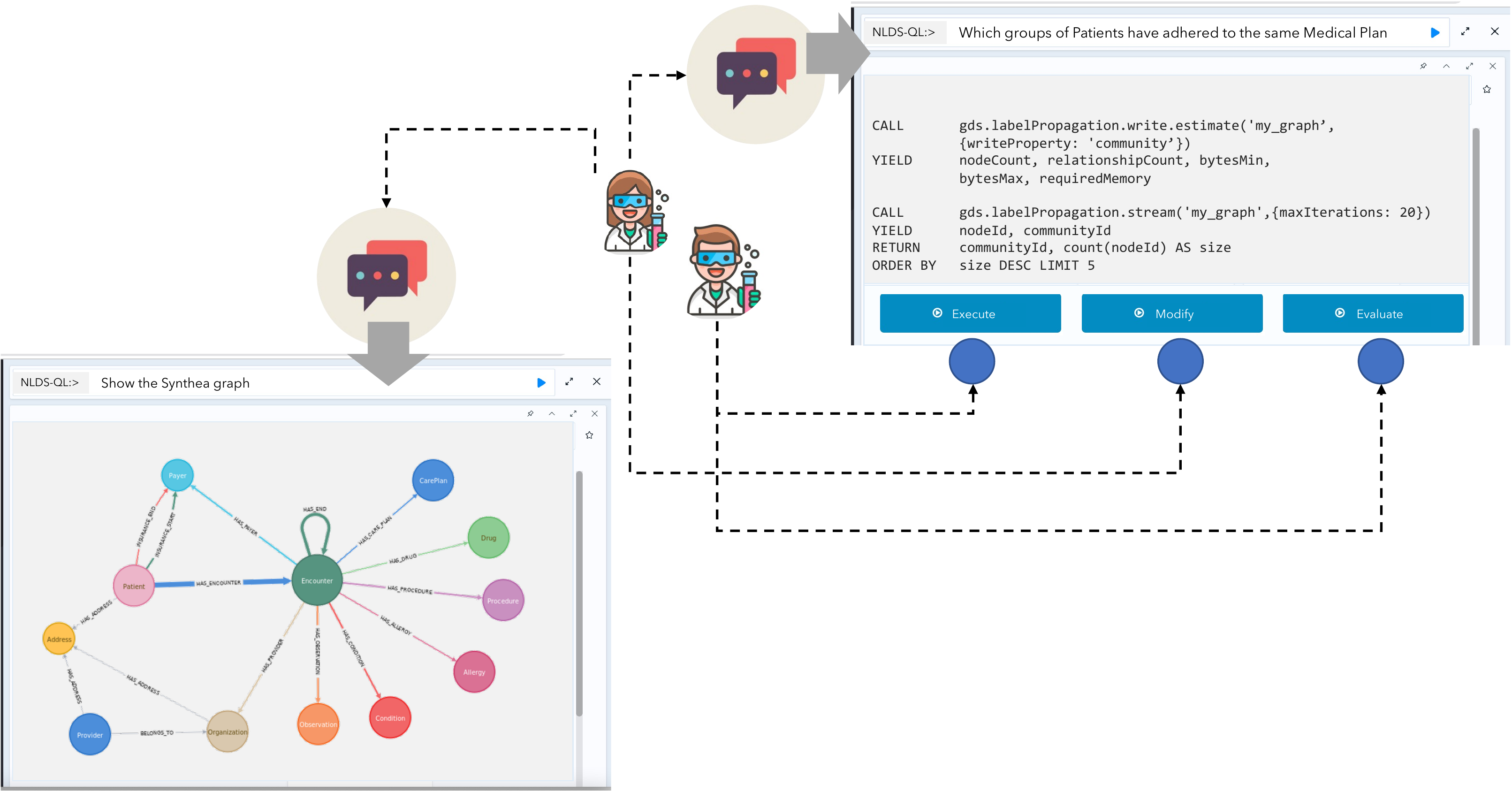}
\caption{Demonstration conversation pipeline
}\label{fig:conversation}
\end{figure*}

The demonstration of NLDS-QL on the Synthea patients graph is based on a conversational pipeline where expert and non-expert users can ask questions to start exploring the graph (see Figure \ref{fig:conversation}). The demonstration environment initially shows the Synthea patients graph, and users can ask for details about the description of the graph, like the number of nodes and relations. Then, the user can ask a question. The system generates one or several queries, and then the user can either choose one or several queries to be adjusted or executed and then modified (see right side of Figure \ref{fig:conversation}). For every choice, the user can evaluate the system's performance with stars that show the degree of satisfaction. 

For exploring the Synthea graph, the demonstration scenario proposes a set of queries that include navigational queries of the type selection, projection, aggregation.

%-----------------------------------------------------------
 %\paragraph
\noindent 
 {\bf Selection.}
 %-----------------------------------------------------------
{\em Find the Medications for which the DESCRIPTION is Lisinopril 10 MG Oral Tablet and the REASON of the DESCRIPTION is Hypertension.}
\begin{verbatim}
    MATCH (n:Allergies) return n.DESCRIPTION
\end{verbatim}
   
%-----------------------------------------------------------
%\paragraph
\noindent 
{\bf Projection.}
%-----------------------------------------------------------
{\em Which is the birthplace of the PATIENTS in the study?}
\begin{verbatim}
    MATCH (n:Patients) return n.BIRTHPLACE 
\end{verbatim}

%-----------------------------------------------------------
%\paragraph
\noindent 
{\bf Selection and Projection.} 
%-----------------------------------------------------------
{\em Find the Encounters DESCRIPTION node where the DESCRIPTION of the drugs is Amlodipine 5 MG Oral Tablet.}
\begin{verbatim}
    MATCH (n:Encounters)-[*]->(m:Medications  
    {DESCRIPTION:'Amlodipine 5 MG Oral Tablet’}) 
    return n.DESCRIPTION, m.DESCRIPTION
\end{verbatim}
  
%-----------------------------------------------------------
%\paragraph
\noindent 
{\bf Aggregation.} 
%-----------------------------------------------------------
{\em How many patients are caucasian?}
\begin{verbatim}
    MATCH (n:Patients {RACE:'white’}) 
    return count(n) 
\end{verbatim}
 
%\end{itemize}

For data science queries, the demonstration shows NLDS-QL  questions that refer to centrality type operations. 
Note that the translation is quite complex as it involves:
\begin{itemize}
    \item Specifying a graph view from the patient graph, as Neo4J works with graph views stored in RAM when data science algorithms are applied.
    \item Then it is possible to generate two queries that call the page rank algorithm to process the keyword "most important" with the possibility to make the view persistent and consider the constraints related to the parameters of the Pagerank algorithm.
\end{itemize}

\noindent
{\bf Centrality.}\\
{\em Find the most popular Encounters for Medications in the graph.}
\begin{verbatim}
    MATCH (n:Encounters)-[r:ENCOUNTER_FOR_MEDICATION]-() 
    with n,count(*) as degree return id(n), degree 
    ORDER BY (degree) DESC
\end{verbatim}

\noindent
{\em Find the most important Drugs prescribed for the
PATIENT with a maximum of 25 iterations and a damping factor of 0.60.}

\begin{verbatim}
CALL gds.graph.create('my_graph','Medications’,
{PATIENT_HAS_MEDICATION: {orientation: 'NATURAL'}}) 

CALL gds.pageRank.write.estimate( 'my_graph', 
        {writeProperty: 'pageRank',
         maxIterations: 25, 
         dampingFactor:0.60}) 
YIELD nodeCount, relationshipCount, bytesMin, 
      bytesMax, requiredMemory 
      
CALL gds.pageRank.stream( 'my_graph' ) 
YIELD       nodeId, score 
RETURN      gds.util.asNode(nodeId).name AS name, score 
ORDER BY    score DESC LIMIT 10
\end{verbatim}

In this example, NSDL-QL generates the template that includes first the graph "my\_graph". Then it computes the estimation of required memory, number of nodes and relations, minimum and maximum bytes that will yield the resulting graph when executing PageRank with the specified parameters. Then the call to the algorithm with the result format with the top 10 nodes associating each node with its score. \\
\noindent
{\bf Community detection.}
The translation also involves several operations as described in the definition of data science queries.

\noindent
{\em Get the subgroup of Patients who have PATIENT\_HAS\_CAREPLAN in the graph with max iterations 20 
}

\begin{verbatim}
CALL gds.labelPropagation.write.estimate('my_graph’, 
        {writeProperty: 'community’}) 
YIELD   nodeCount, relationshipCount, bytesMin, 
        bytesMax, requiredMemory 

CALL gds.labelPropagation.stream('my_graph',
          {maxIterations: 20}) 
YIELD     nodeId, communityId 
RETURN    communityId, count(nodeId) AS size 
ORDER BY  size DESC LIMIT 5

\end{verbatim}
%________________________________________________________
\section{Conclusion and Results}\label{sec:conclusion}
%________________________________________________________
The demonstration of the  evaluator NLDS-QL  
shows how to map NL data science questions (using an adapted vocabulary) to Neo4J data science query templates. 
 The demonstration is based on a use case on querying and analysing a graph in the medical domain. Users with medical and non-medical backgrounds can define a sequence of natural language queries executed step by step to explore the graph, as in data science questions. Thereby users can acquire an understanding of medical prescriptions proposed to patients by classifying 
their treatment, their physiological characteristics to better understand how diseases are diagnosed and treated according to patients conditions.
In this way, we 
show the essential aspects of a data science query template expressed in NL.

The approach is flexible and can be enhanced for processing documents with richer NL vocabulary and more complex templates.
 The intervention of a human in handling natural language queries calls for the design of an interactive strategy based on conversation. We have started to design a more evolved conversational interface considering human in the loop and user profiling techniques.

\bibliographystyle{ACM-Reference-Format}
\bibliography{sample-base} 

\end{document}